\documentclass{ws-procs9x6}
\newif\ifpdf \ifx\pdfoutput\undefined \pdffalse \else \pdftrue \fi 
\usepackage{amsmath}
\usepackage{amssymb}
\usepackage{fancyhdr}
\usepackage{journals}
\usepackage{color}
\ifpdf
 \usepackage[colorlinks,hyperindex]{hyperref}
\else
 \newcommand{\href}[2]{#2}                   
 \usepackage[dvips,pagebackref]{hyperref}    
\fi 
\renewcommand\topmargin{0in}

\def \Vista {{\sc{Vista}}}
\def \Sleuth {{\sc{Sleuth}}}
\def \Quaero {{\sc{Quaero}}}
\def \Bard {{\sc{Bard}}}

\begin{document}

\title{Solution to the LHC Inverse Problem}

\author{Bruce Knuteson}

\address{MIT \\
knuteson@mit.edu}

\maketitle

\abstracts{The ``LHC Inverse Problem'' refers to the question of determining the underlying physical theory giving rise to the signals expected to be seen at the Large Hadron Collider.  The solution to this problem (\Bard) is reviewed.  The combination of CDF and D\O\ data is motivated.}

These proceedings explore the incorrectness of each of the three words in ``LHC Inverse Problem.''  We work backward.  Sections I, II, and III discuss ``Problem,'' ``Inverse,'' and ``LHC,'' respectively.

\section{Problem}

The LHC Inverse Problem is fortunately no longer a problem, since it has a solution in the form of an algorithm named Bard\cite{BardPRL:Knuteson:2006ha}.  \Bard, named after Shakespeare, systematically weaves a set of stories to explain a particular discrepancy observed in the data.

Unfortunately no proof exists that \Bard\ is the correct solution, the best solution, or even a good solution.  The best solution to the LHC Inverse Problem will be the one leading most quickly to an underlying theory given the particular discrepancies between data and Standard Model prediction observed at the LHC.  Which solution is best thus depends on how Nature actually behaves, which is what are trying to figure out. 

The claim that \Bard\ is the solution to the LHC Inverse Problem is simply the statement that \Bard\ is the only systematic and reasonably encompassing solution to the problem so far proposed.  \Bard\ is an experimentalist's solution to a theorist's problem, starting directly from a specific discrepancy observed in data, and working with diagrammatic explanations that are easily visualized and understood.  The essence of the algorithm is conveyed in a single sentence: Systematically draw all conceivable tree-level Feynman diagrams, introducing new particles and couplings as necessary, where the incoming legs of the diagram are determined by the colliding particles, and the outgoing legs are determined by the events in which the signal is seen.  Getting to the right answer quickly requires making reasonable assumptions to simplify the exploration of explanation space.  \Bard\ provides a convenient framework for imposing specific assumptions to reduce the size of this space.

For \Bard\ (or any other proposed solution to the LHC Inverse Problem) to work in practice, a means of quickly, robustly, efficiently, and automatically testing any arbitrary hypothesis against the data is required.  \Quaero\cite{QuaeroPRL:Abazov:2001ny} is the solution to this problem.

If we need \Bard\ to tell us what we have found, then we clearly did not know what we were looking for in the first place, so how did we find it?  Some means of performing a quasi-model-independent search for new physics is required.  \Sleuth\cite{SleuthPRL:Abbott:2001ke,SleuthPRD1:Abbott:2000fb,SleuthPRD2:Abbott:2000gx} or a General Search\cite{H1GeneralSearch:Aktas:2004pz} is the solution to this problem.

Before performing a quasi-model-independent search for new physics on the high-$p_T$ tails, an understanding of the gross features of the entirety of the high-$p_T$ data is desired.  \Vista\cite{ACAT2003:Knuteson:2004nj} is the solution to this problem.  \Vista\ could play an important role in the physics commissioning of the LHC experiments.

The phrase ``the solution'' used throughout this section unfortunately does not mean these algorithms are better than their competition, but simply that there is no competition.

\section{Inverse}

The LHC Inverse Problem is an ``inverse'' problem if you are used to working out the phenomenological consequences of a particular new physics scenario; inferring the new physics scenario from the phenomenological consequences is then backward.  The problem is thus an inverse problem only if you are a theorist.  If you are an experimentalist, it is just another problem.

The naming of this ``Inverse'' problem implies the forward direction involves going from a model to its phenomenological consequences.  Indeed, most searches for new physics in our field start with a particular model, work out its consequences, and then fail to observe these consequences.  The scientific method taught in grade school has it the other way around:  make measurements on a poorly understood system, observe unexplained phenomena, and build a theory to explain existing observation and predict future observation.  The grade school version is much more exciting.

\section{LHC}

The LHC Inverse Problem is an ``LHC'' problem because few in our field believe a surprising discovery will be made at the Fermilab Tevatron.  

We have the potential for leaving a negative Tevatron legacy.  It is not unlikely that after discrepancies have begun to be sorted out at the LHC it will be widely recognized by the community that a discovery could have been made at the Tevatron.  This will be a tragedy and a missed opportunity.  The Tevatron community should do whatever it can at small cost to ensure this does not happen.

The {\em{de facto}} discovery threshold in our field has become ``$5\sigma$,'' with the discovery of the top quark frequently pointed to as the justification for this choice of threshold.  The top quark discovery was announced when a statistical significance of $4.8\sigma$ had been achieved at CDF\cite{CDFTopDiscovery:Abe:1995hr} and $4.6\sigma$ had been achieved at D\O\cite{D0TopDiscovery:Abachi:1995iq}.  From the point of view of someone outside Batavia, Illinois, the top quark was a Tevatron discovery at a statistical significance of $4.6\sigma + 4.8\sigma = 7\sigma$.

The Tevatron experiments are showing results on 1~fb$^{-1}$ of data at this year's Summer conferences.  The total integrated luminosity accumulated by the Tevatron experiments, with all good run lists applied, is twice this:  1~fb$^{-1}$ (CDF) + 1~fb$^{-1}$ (D\O) = 2~fb$^{-1}$.  

The Tevatron integrated luminosity is roughly doubling each year.  It follows that combining CDF and D\O\ data will lead to a discovery one year earlier than not combining these data.  With LHC turn on drawing near, there is little time to waste.  

The combination of published results is not being advocated, but rather combination of the data in the exploration phase.  A global analysis of CDF data should be performed, systematically searching all CDF high-$p_T$ data for discrepancies with Standard Model prediction.  In parallel, a global analysis of D\O\ data should be performed, systematically searching all D\O\ high-$p_T$ data for discrepancies with Standard Model prediction.  The global analyses of CDF and D\O\ data should be conducted within a common framework so that histograms can be filled containing both CDF and D\O\ events, with the overlaid Standard Model prediction the sum of the Standard Model prediction at D\O\ and the Standard Model prediction at CDF.  \Vista\ and \Sleuth\ provide a natural framework for this effort.

Simultaneous analysis of CDF and D\O\ data will converge more quickly than the analysis of either alone.  When a discrepancy between data and Standard Model prediction is observed in one experiment, the debugging process must address whether the discrepancy is due to a detector effect or a poor implementation of the Standard Model prediction.  On a single experiment, the debugging effort is split between these two possibilities.  In a combined analysis, the presence of the effect in the other experiment focuses the debugging effort on a poor implementation of the Standard Model prediction; the absence of the effect focuses the debugging effort on a possible detector problem.  A combined Tevatron analysis can be achieved within six months, if the two collaborations want this to occur.  

Maximizing the chance of a Tevatron discovery requires a combined global high-$p_T$ analysis at CDF and D\O.

\vspace{0.1in}

These proceedings have reviewed the LHC Inverse Problem.  It is argued that this problem now has a solution, in the form of an algorithm named \Bard; that this problem is an inverse problem only from the point of view of those working top down; and that this is a Tevatron problem until the LHC turns on.  The Fermilab Tevatron will see at least three more doublings of its data: from 1~fb$^{-1}$ per experiment to 2~fb$^{-1}$; from 2~fb$^{-1}$ to 4~fb$^{-1}$; and a doubling as soon as a combined global analysis is achieved at CDF and D\O.  Whether we can fully capitalize on this opportunity in the next two years remains to be seen.

\bibliography{dis06}

\end{document}